\documentclass[reprint,10pt,secnumarabic,amssymb, nobibnotes, aps, prl,groupedaddress,superscriptaddress]{revtex4-1}
\usepackage{graphicx}
\usepackage{dcolumn}
\usepackage{bm}
\usepackage{siunitx}
\usepackage[dvipsnames]{xcolor}
\usepackage{lipsum}
\usepackage{verbatim}
\usepackage{soul}
\usepackage{amsmath}
\usepackage{comment}
\usepackage{setspace}
\usepackage{pdfpages}
\usepackage{etoolbox} 
\usepackage{xurl}
\usepackage{multirow}

\makeatletter                                                                                    
\patchcmd{\@outputpage@head}{\@ifx{\LS@rot\@undefined}{}{\LS@rot}}{}{}{}
\makeatother

\begin{document}
\title{Plasma Accelerator For Decaying Particles}
\author{Chiara Badiali}
\affiliation{GoLP/Instituto de Plasmas e Fus\~ao Nuclear, Instituto Superior T\'ecnico, Universidade de Lisboa, 1049-001 Lisbon, Portugal}

\author{Rafael Almeida}
\affiliation{GoLP/Instituto de Plasmas e Fus\~ao Nuclear, Instituto Superior T\'ecnico, Universidade de Lisboa, 1049-001 Lisbon, Portugal}

\author{Bernardo Malaca}
\affiliation{GoLP/Instituto de Plasmas e Fus\~ao Nuclear, Instituto Superior T\'ecnico, Universidade de Lisboa, 1049-001 Lisbon, Portugal}
\affiliation{Centro Nacional de Computação Avançada / Deucalion Supercomputer, Portugal}

\author{Ricardo Fonseca}
\affiliation{GoLP/Instituto de Plasmas e Fus\~ao Nuclear, Instituto Superior T\'ecnico, Universidade de Lisboa, 1049-001 Lisbon, Portugal}
\affiliation{DCTI/ISCTE, Instituto Universitário de Lisboa, 1649-026, Lisboa, Portugal}

\author{Thales Silva}
\affiliation{GoLP/Instituto de Plasmas e Fus\~ao Nuclear, Instituto Superior T\'ecnico, Universidade de Lisboa, 1049-001 Lisbon, Portugal}

\author{Jorge Vieira}
\affiliation{GoLP/Instituto de Plasmas e Fus\~ao Nuclear, Instituto Superior T\'ecnico, Universidade de Lisboa, 1049-001 Lisbon, Portugal}
\date{\today}
\begin{abstract}
We introduce a plasma wakefield acceleration scheme capable of boosting initially subrelativistic particles to relativistic velocities within millimeter-scale distances. A subluminal light pulse drives a wake whose velocity is continuously matched to the beam speed through a tailored plasma density, thereby extending the dephasing length. We develop a theoretical model that is generalizable across particle mass, initial velocity, and the particular accelerating bucket being used, and we verify its accuracy with particle-in-cell simulations using laser drivers with energies in the Joule range.
\end{abstract}
\maketitle

\textit{Introduction}---The particle physics community is actively considering muon colliders as one possible next step for high-energy physics studies~\cite{Long2021}. Muons, like electrons, are fundamental particles, meaning their total energy is available in collisions, unlike hadrons~\cite{Costantini2020,Chiesa2020,Zimmermann2018}. Nonetheless, due to their finite lifetime ($\tau_\mu = \SI{2.2}{\micro\second}$ in the proper frame), muons must be collected, cooled, and rapidly accelerated before a significant number of them decay~\cite{MICE2020}, which is a challenging task for conventional accelerators that use radio-frequency (RF) cavities to accelerate particles. Similarly, pions and kaons, which have even shorter decay times ($\tau_{\pi^{\pm}}= \SI{26}{\nano\second}$ and $\tau_{k^{\pm}} =\SI{12}{\nano\second}$ in the proper frame), are nearly impossible to accelerate efficiently with standard RF accelerators \cite{Apyan2023}.
Plasma acceleration, with its $\sim \SI{}{GV/cm}$
gradients \cite{arjmand2023shot,Modena1995,Hogan2005,Malka2012}, can boost short-lived beams to relativistic energies before they decay. 

The fraction of particles surviving an acceleration process can be estimated as $S\simeq(\gamma_i/\gamma_f)^{\nu_p}$ \cite{GoldmanSilbar2008,Noble1992}, with $\gamma_i$ and $\gamma_f$ being the initial and final Lorentz factor, and $\nu_p =(m_pc)/(qE_{acc}\tau_p)$—where $m_p$ is the particle's rest-mass, $c$ the speed of light in vacuum, $E_{acc}$ is the accelerating field, $\tau_p$ the proper lifetime, and $q$ the particle's charge. For a representative $\mu^{-}$ bunch accelerated from \SI{200}{MeV} to \SI{1}{GeV} in a \SI{0.2}{GV/cm} plasma stage, one expects a loss of $\sim 10^{-6}$ of the initial particles, whereas a \SI{20}{MV/m} RF linac yields to 1 $\%$ particles' loss.
For $\pi^{-}$ the contrast is stronger: a fraction of $\sim 10^{-3}$  particles decay in a plasma-based accelerator, in contrast to 94 $\%$ for the RF.

Kaons, pions, and muons are often created with subrelativistic velocities due to their large masses. A possible solution to accelerate these particles is to employ accelerating structures with a subluminal phase-velocity~\cite{peano2009prospects,Pukhov2023}. One of the first configurations relied on direct laser acceleration methods, with subluminal phase-speeds, using two counterpropagating laser beams, have been proposed for muon acceleration~\cite{peano2009prospects}. Here, we propose a plasma wakefield acceleration method working in a co-propagating geometry. 



The use of plasma wakefields~\cite{DeLaOssa2013,Costa2022} to accelerate subrelativistic particles requires drivers traveling below the speed of light. These drivers are now available due to recent advances in ultra-fast optical shaping techniques~\cite{Kondakci2019,Froula2018,SainteMarie2017}. One example is space-time wave-packets~\cite{yessenov2022space}, which use pre-engineered correlations between wave vectors and frequency to yield shape-invariant electromagnetic wave-packets whose intensity peak can propagate with arbitrary velocity. The distance over which these features persist depends on the laser energy~\cite{almeida2025}. Although subluminal drivers can excite wakefields that trap non-relativistic particles, energy gain is ultimately limited by dephasing as the beam eventually overtakes the slow driver~\cite{Sadler2020,Lu2007} before reaching relativistic velocities.


 In this Letter, we circumvent these limitations by introducing a versatile subluminal-pulse-driven plasma accelerator that, together with a carefully tailored density profile~\cite{Pukhov2008,Rittershofer2010,Xu2017}, efficiently boosts initially subrelativistic particles to relativistic speeds. 
 At that point, traditional wakefield drivers can further boost their energies~\cite{Geng2024}. %
The proposed mechanism achieves rapid acceleration across both linear and nonlinear wakefield regimes. We validate our theoretical predictions with particle-in-cell (PIC) simulations with the code OSIRIS~\cite{2002Fonseca}, demonstrating how muons, pions, and other particles reach relativistic energies within millimeter-scale distances. Our findings suggest that our subluminal-accelerator concept can be used with heavier hadrons, like kaons and protons, and may open new possibilities for compact plasma-based accelerators in various high-energy and applied-physics contexts. In the linear regime, the theoretical model extends naturally to positive particles, like anti-muons and anti-pions. We illustrate the concept considering laser energies in the Joules range that have already been demonstrated for pulses with spatio-temporal couplings~\cite{Liberman2024}.

\begin{figure}[b]
  \centering \includegraphics[width=1.\linewidth]{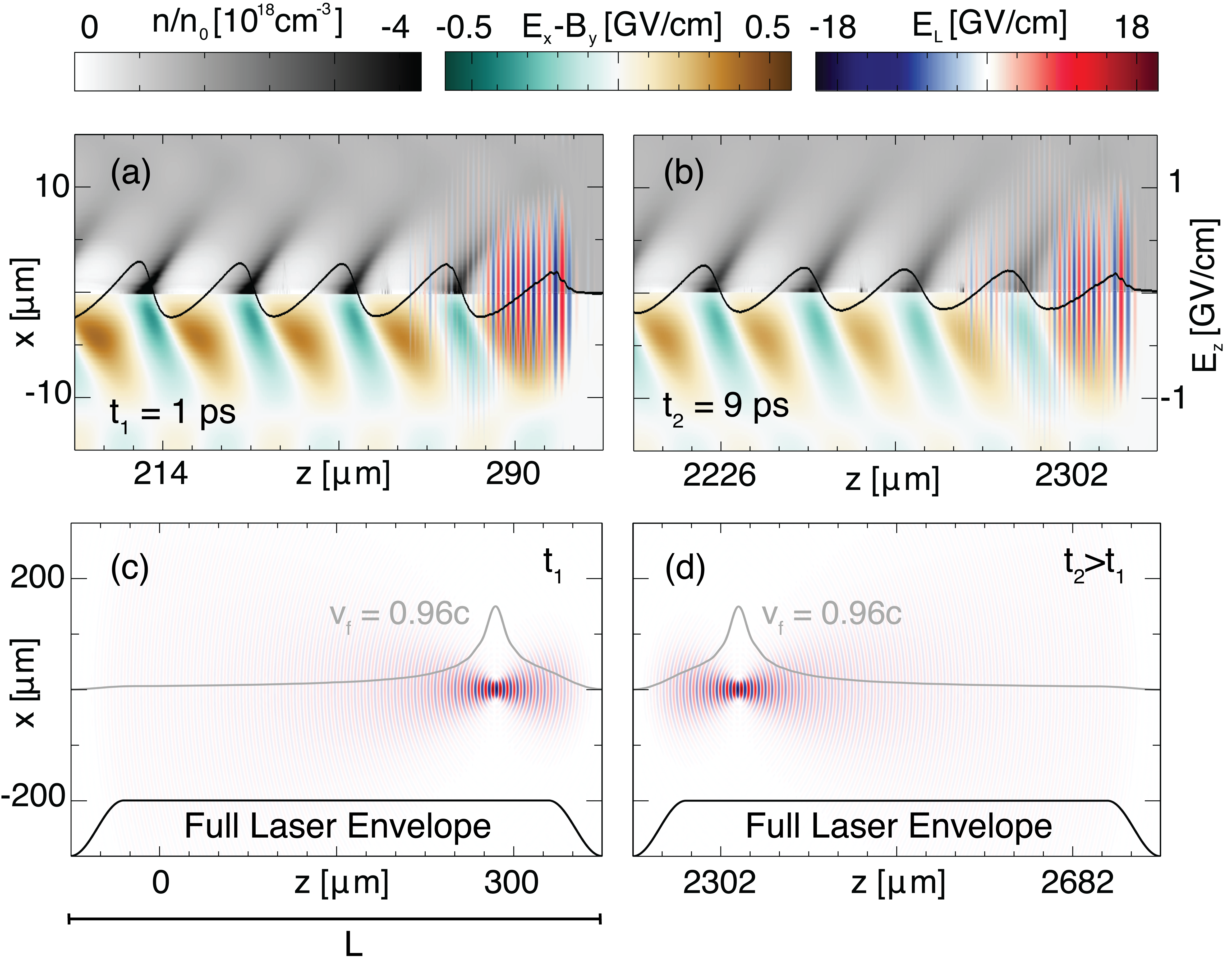}
  \caption{OSIRIS simulation of a subluminal ($v_f=0.96c$) spatio-temporal driver pulse in plasma at (a) $t_1=\SI{1}{\pico\second}$ and (b) $t_2=\SI{9}{\pico\second}$. Key features include plasma density (gray), transverse field $E_x-B_y$ (green/orange), on-axis accelerating field $E_z$ (black line), and the laser pulse (blue/red). The wakes are stable, conserving the acceleration structure. In panels (c) and (d), for the same simulation, we show the full laser envelope (black) and the localized intensity peak (gray) that generates the wake (blue–red). Because the envelope travels at $c$ while the focus advances at $v_f<c$, the peak slips backward within the envelope between the two times (c) $t_1$ and (d) $t_2>t_1$. Once the peak reaches the envelope tail, the pulse no longer drives efficient wakes.
  }
  \label{fig:1}
\end{figure}

\textit{Setup and Analytical Model}---To investigate the wake excitation driven by a space-time wave packet, we start by showing a simulation of a pulse whose focal velocity is $v_f = 0.96\,c$, central wavelength $\lambda = \SI{0.8}{\micro\meter}$, normalized vector potential of $a_0 \equiv eE_L/m_ec\omega_0 = 3$ as it propagates through a uniform plasma of density $n_0 = 10^{18}\mathrm{cm^{-3}}$, resulting in $\omega_0/\omega_p =20$, where $\omega_0$ is the laser frequency, $\omega_p = (n_0e^2/m_e\epsilon_0)^{1/2}$ is the plasma frequency, $e$ is the elementary charge, $m_e$ the electron mass, $\epsilon_0$ is the vacuum permittivity, and $E_L$ is the amplitude of oscillation of the laser electric field. All the simulations in here were performed using azimuthal-mode-decomposed cylindrical coordinates~\cite{lifschitz2009particle,davidson2015implementation}; simulation details are present in the Appendix. Figures \ref{fig:1}(a) and \ref{fig:1}(b) summarize the key wake features: (i) a longitudinal accelerating field $E_z \simeq \SI{0.2}{GV/\centi\meter}$, (ii) alternating focusing/defocusing transverse fields $E_x-B_y$, and (iii) an accelerating structure that remains stable over the simulated interval.  

The structure of the spatio-temporal laser pulse is further detailed in Figs.~\ref{fig:1}(c) and (d). These panels illustrate how the pulse's peak intensity (gray line), having a length on the order of $\sim\SI{20}{\micro\meter}$, travels at a subluminal focus velocity $v_f < c$, while the full laser envelope (black line), with a length on the order of $L\sim \SI{500}{\micro\meter}$, propagates at $v_{env}\sim c$. The full pulse length L and the velocity difference between the full envelope and the focal velocities determine the duration $\Delta t \simeq L / (v_{env} - v_f)$ over which the structured pulse can sustain effective acceleration.

Although slow beams can be phase-locked within the wakes generated by these subluminal pulses, the effective acceleration length in a uniform plasma is ultimately limited by dephasing~\cite{sadler2020overcoming}.
To prolong the dephasing length, we employ a tailored plasma density profile, as plasma density gradients can provide an additional degree of freedom to control the phase velocity \cite{Pukhov2008, rittershofer2010tapered,PhysRevAccelBeams.20.111303}. It is then possible to obtain phase-locking over extended distances by matching this changing wake velocity to the evolving speed of the accelerated beam. 

\begin{figure}[t]
  \centering
  \includegraphics[width=1.0\linewidth]{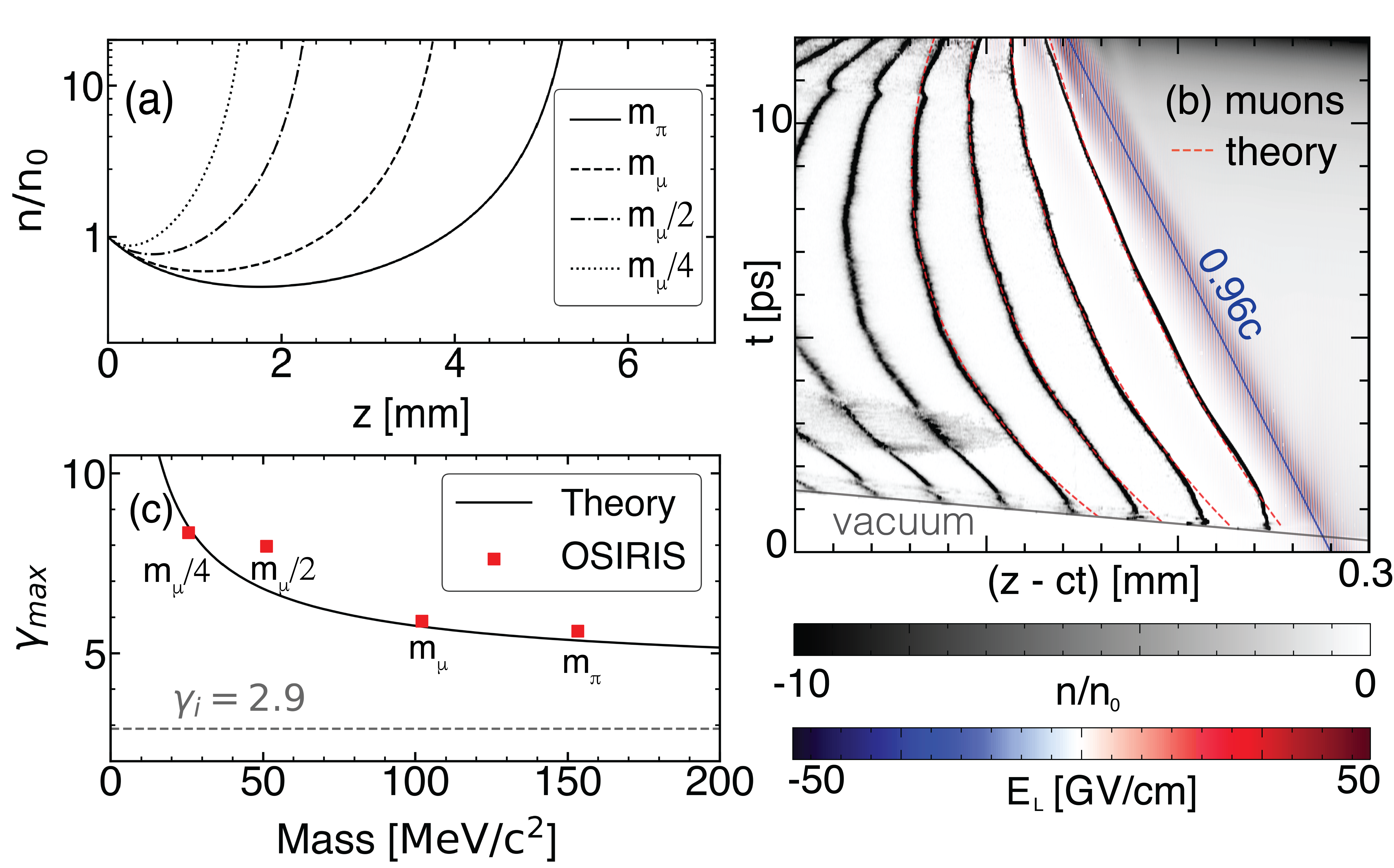}
  \caption{\label{fig:epsart} Shown here is an overview of our subrelativistic acceleration scheme for muons, pions, and artificially scaled muon masses, all initially at $0.94\,c$ and driven by a driver moving at $0.96\,c$:
(a) Tailored plasma density profiles were used in these simulations, starting with a decreasing density region to capture the subrelativistic beams, followed by an up-ramp that extends phase-lock and acceleration. 
(b) Waterfall plot of the driver and plasma response for muon acceleration, revealing how the density ramps (grayscale) slow down or speed up the wake while the structured pulse (in red/blue) maintains its subluminal velocity at $0.96\,c$. This combination of fixed driver speed and density shaping allows muons to transition smoothly from $0.94\,c$ to higher energies within a single stage.
(c) Final maximum $\gamma$ factor (energy) for each particle beam; solid lines are theoretical predictions, and red squares are OSIRIS simulation results.}
  \label{fig:2}
\end{figure}

We can determine the ideal density profile by noting that the accelerating electric field is proportional to the local plasma density $E_{z} \propto \sqrt{n(z)}$~\cite{esarey1996overview} and imposing that the particle local velocity matches the wake's phase velocity generated by the subluminal driver. We then obtain a system of coupled equations (more details in Appendix 1)
\begin{subequations}
\begin{gather}
\frac{d[\gamma(z) v(z)]}{dz} = \alpha_0 \frac{q}{e}  \frac{m_e}{m_p}\frac{c}{v_f} \omega_{p0} \sqrt{\frac{n(z)}{n_0}}, \label{eq:1a}\\
v = v_f - 
\frac{2 \pi n_w v_f^2}{\omega_{p0}}\frac{d}{dz} \sqrt{\frac{n_0}{n(z)}},\label{eq:1b}
\end{gather}
\label{eq:denwake}
\end{subequations}
where $v(z)$ is the accelerated particle velocity, $\gamma\approx(1-v(z)^2/c^2)^{-1/2}$ is the Lorentz factor, $n_0$ is a reference plasma density (here, we use as the initial plasma density of the ideal profile), $\omega_{p0}$ is the plasma frequency for the $n_0$ density, $\alpha_0$ is the electric field magnitude in units of $m_ec\omega_{p0}/e$, $q$ is the charge of the particle, and $n_w$ is the number of the trailing wake where the particle is accelerated. 
Solving Eq.~\eqref{eq:denwake} yields the ideal, tailored plasma density profile $n(z)$ that preserves phase-locking. 
Figure ~\ref{fig:2}(a) shows numerical solutions of Eq.~\eqref{eq:denwake} considering the first wake of acceleration ($n_w=1$) for beams of different particle masses, each with an initial velocity of $0.94\,c$ and a driver velocity of $0.96\,c$. They share a similar structure: an initial downward slope decelerates the back of the accelerating bucket, allowing the beam to be trapped and begin accelerating. Since the driver velocity is slightly higher than the beam's, the slope can remain gentle, mitigating strong electron self-injection and preventing any substantial reduction of the accelerating field. Next, to prevent dephasing as the beam velocity reaches the driver's focal speed, Eq.~\eqref{eq:denwake} predicts a region of increasing plasma density that decreases the wakefield wavelength and maintains phase-locking. Eventually, the process will be limited by the beam overtaking the driver.
\newline 
\textit{PIC Validation and Scaling}---The proposed theory has been tested through self-consistent PIC simulations. Considering the ideal plasma density profile for the muon beam [Fig. \ref{fig:2}(a)], Fig. \ref{fig:2}(b) shows a waterfall plot of the plasma density and the laser driver electric field as a function of time, along the propagation direction on axis. The sequence of alternating lighter and darker regions behind the driver represents electron density void and accumulation. 
We notice that the trajectory associated with the regions where electrons accumulate slows down in down-ramp regions and speeds up in up-ramp regions, while the structured light pulse focus propagates at a nearly constant velocity. Figure \ref{fig:2}(b) also shows that the phase velocity of the acceleration buckets depends on their distance relative to the driver.  These bucket-velocity changes are smaller (higher) for buckets closer (farther) from the driver. This suggests that buckets farther behind the driver could be beneficial to trap slow particles \cite{Pukhov2023}.

Figure \ref{fig:2}(c) shows a comparison of the maximum particle energy obtained from simulations and theoretical predictions as a function of the particle mass. The theoretical line is calculated using Eq. \eqref{eq:denwake} by estimating the dephasing time as the time it takes for the beam to reach the laser intensity peak.
We perform simulations for particles of different masses, always injecting the beam particles with the same initial velocity $0.94\,c$, behind a laser driver whose intensity peak travels slightly faster at a focal velocity $v_f = 0.96\,c$. For each particle mass, the plasma is initialized using the optimal density profile calculated from Eq. \eqref{eq:denwake}. We observe a good agreement between the theoretical line and the simulation results, demonstrating that our concept is robust for a wide range of particle masses. 
\begin{figure}[b]
  \centering
  \includegraphics[width=1.0\linewidth]{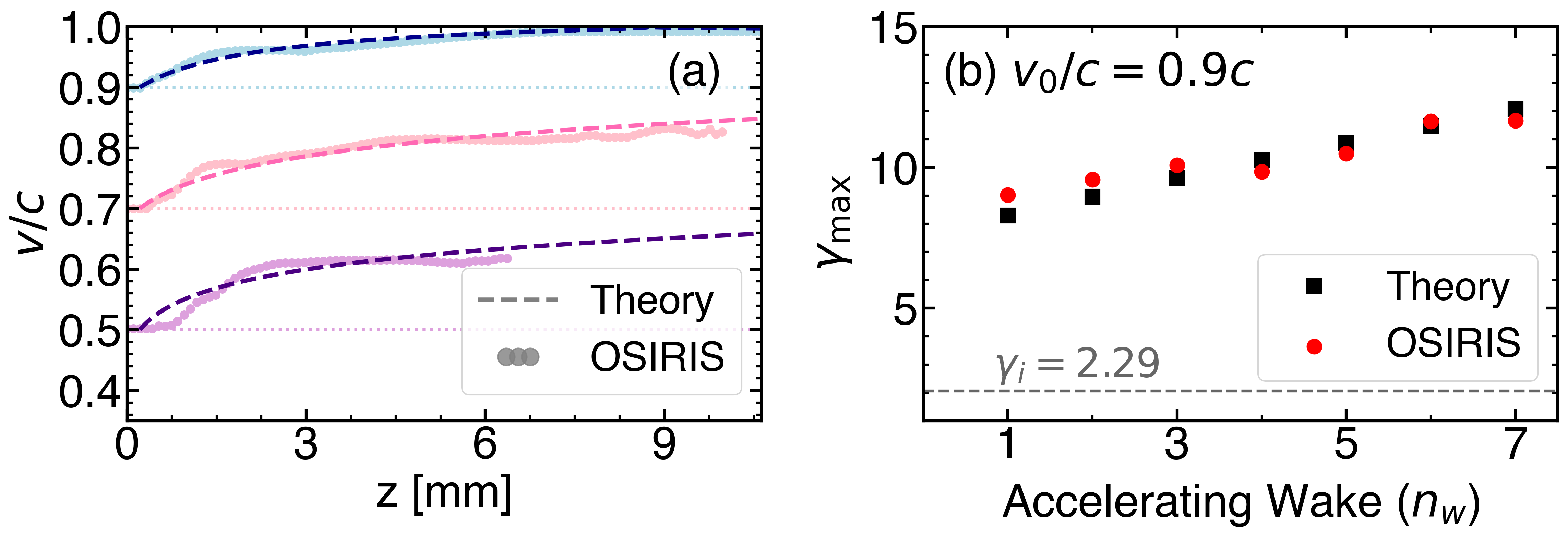}
  \caption{(a) Maximum velocity evolution $v/c$ of three beams starting at $v_0/c = 0.5$, $v_0/c = 0.7$, and $v_0/c = 0.9$. Each beam is driven with focus velocities of $0.8c$, $0.9c$, and $0.97c$, respectively. The dashed lines represent the theoretical velocity evolution, and the solid markers are from OSIRIS simulations. (b) Maximum energy $\gamma$ reached by a $m_\mu/4$ beam initially at $0.9c$, accelerated in different wakes by a driver at $0.97c$. The black squares are theoretical predictions from our density-tailoring model, and the red circles are OSIRIS simulation results. }
  \label{fig:3}
\end{figure}
The simulation parameters used to perform these simulations are summarized in the Appendix 2. 

We also explored the acceleration of beams with different initial velocities. For beam particles with $m_p = m_\mu/4$, we performed simulations with initial beam velocities of $0.5\,c$, $0.7\,c$, and $0.9\,c$, with a corresponding driver velocity of $0.8\,c$, $0.9\,c$, and $0.97\,c$, respectively. In Fig. \ref{fig:3}(a), the dashed lines show the velocity evolution $(v/c)$ predicted by our model [Eq. \eqref{eq:denwake}], while the solid markers are the maximum particle velocity from the PIC simulations, resulting in a good agreement for all cases. 


Equation \eqref{eq:denwake} also predicts the ideal plasma profile as a function of the bucket position $n_w$. Figure \ref{fig:3}(b) shows the final $\gamma$ factor reached by beams accelerated in the first seven buckets behind the driver, considering $m_p = m_\mu/4$. We use the same configuration as for the blue curve in Fig. \ref{fig:3}(a) (beam initially moving at $0.9\,c$ and the driver at $0.97\,c$). 
Because a later plasma bucket reacts more strongly to a given density variation than the first, its phase velocity also drops faster than that of the first bucket in a region of decreasing plasma density. Thus, we can use a gentle down-ramp to keep the beam in phase with the accelerating bucket, maintaining a high amplitude of the accelerating field. Although the initial field in a secondary bucket is slightly weaker than that of the first, the mild taper is able to maintain the accelerating field. Hence, the beam ultimately gains more energy. 

\begin{figure}[b]
  \centering
  \includegraphics[width=1.\linewidth]{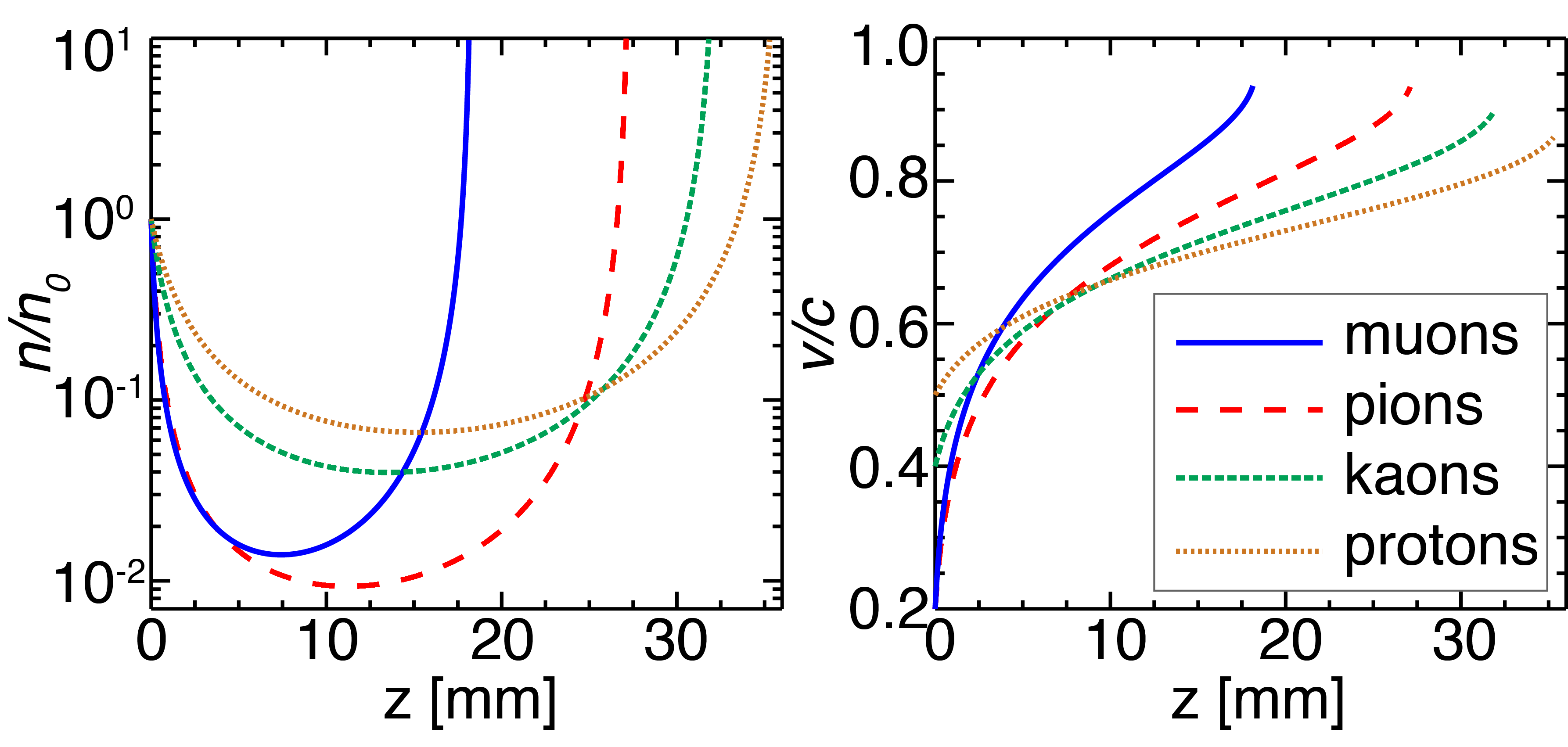}
  \caption{Density profile (a) and theoretical velocity gain (b) for the acceleration of various particles from very low energies (0.2c for muons, 0.2 for pions, 0.4c for kaons and 0.5c for protons) using a laser driver with $a_0 = 0.8$ and $\omega_0/\omega_p = 20$. }
  \label{fig:4}
\end{figure}
To explore even more demanding scenarios, we extended our theoretical model in Eq. \eqref{eq:denwake} to study the acceleration of heavier hadrons. Figure ~\ref{fig:4}(a) illustrates the predicted ideal density profile and velocity evolution for these more extreme cases. Figure (b) shows that muons, pions, kaons, and protons can be accelerated in tens of millimeters from $v\le 0.5\,c$ to $v\sim 0.9\,c$.

The required laser parameters necessary for the acceleration of these heavier particles are within reach. To show this, we combine an estimate for laser energy with an estimate for the particle energy gain. To determine the laser energy, we first note that the spatio-temporal pulses used here have a transverse Gaussian profile where the full length of the longitudinal envelope is $L$ [see Fig.~\ref{fig:1}(c) and (d)]. Thus, the pulse power $P$ is constant along the pulse, and the pulse energy is  $\varepsilon = P\,L/c = \pi w_0^2 I_0 L/c$~\cite{saleh1991chapter3}, where $I_0 = c\epsilon_0 E_0^2/2$ is the peak intensity of the laser. One can write an engineering formula for the laser energy as
\begin{equation}
\varepsilon\,[J]\approx0.1\,a_0^2\, w_0^2[\SI{}{\micro m}]\, \lambda^{-2}[\SI{}{\micro m}]\, L[\SI{}{\milli\meter}].
\label{eq:ene_1}
\end{equation}
The total length of the laser envelope $L$ sets the propagation distance over which the localized intensity peak persists within the pulse [see Fig. \ref{fig:1}(c) and (d)]. The time it takes for the intensity peak to slip backwards from the front to the back of the full beam envelope is given by $\Delta t \approx L / (c - v_f)$. Thus, the peak intensity travels a total length $\Delta l\approx\Delta t\, v_f = L\,v_f/(c - v_f)$. These spatio-temporal scales can be seen as analogous to pump depletion for space-time wavepackets. 

The energy gained by a particle beam in a spatio-temporal pulse acceleration stage can be estimated with
$\Delta W = m_p c^2\Delta\gamma=F \Delta l = q\,\langle E_{acc}\rangle\,
{L\,v_f}/({c - v_f}),$
where $\langle E_{acc}\rangle$ is the average accelerating field. We can rewrite the length $L$ in terms of the energy gain $\Delta\gamma$, focal velocity, particle charge, and $\langle E_{acc}\rangle$ and replace in Eq. \eqref{eq:ene_1}, resulting in
\begin{equation}
\varepsilon\, [\SI{}{J]}\approx\frac{a_0^2\Delta\gamma}{20} \frac{e}{q}\frac{m_p}{m_e}\left(\frac{c}{v_f}-1\right){E_{acc}^{-1} \left[\SI{}{\frac{GV}{m}}\right]}\left(\frac{w_0}{\lambda}\right)^2.
\label{eq:ene_2}
\end{equation}
With the parameters $a_0 = 0.8$, $w_0=\SI{3.6}{\micro m}$, $\lambda = \SI{0.8}{\micro m}$, $v_f = 0.97\,c$, and $E_{acc}= \SI{20}{GV/m}$, the required laser energy to obtain an energy gain of $\Delta \gamma = 5$ is $\SI{1}{J}$, $\SI{1.3}{J}$, $\SI{4.9}{J}$ and $\SI{9.2}{J}$ for muons, pions, kaons, and protons, respectively. These energy requirements are within the capabilities of state-of-the-art high-intensity laser systems used in advanced laser-plasma experiments \cite{Liberman2024}, supporting the experimental feasibility of this scheme.

\textit{Conclusions}---We have shown that subluminal light pulses combined with carefully tailored density profiles can accelerate initially subrelativistic decaying particles (\textit{e.g.}, muons and pions) to relativistic energies over remarkably short distances. We have developed a model to calculate the ideal plasma profile by matching the wake’s velocity to the evolving speed of the beam. Our results were validated through numerical simulations, resulting in an excellent agreement in predicting the beam energy gain. Because the scheme operates with few-joule drivers, which have already been demonstrated in spatio-temporal optics experiments~\cite{Liberman2024}, exploring plasma-based acceleration of decaying particles is within reach, and it is an application of plasma acceleration that might not have an analog in standard RF accelerators. This approach may serve as a rapid booster stage for decaying particles in muon-collider designs or as a compact source of high-energy pions, kaons, and muons for neutrino physics, opening a range of scientific and societal applications. The versatility of the scheme could allow for its application to improve electron acceleration in plasma wakefields and isotope separation.
\begin{acknowledgments}
We acknowledge Luis O. Silva and Pablo Bilbao for fruitful discussions. We gratefully acknowledge EuroHPC for awarding us access to LUMI-C at CSC
(Finland). This work was supported by FCT I.P. under Project 2024.07895.CPCA.A3 – DOI: https://doi.org/10.54499/2024.06987.CPCA.A3 – at MareNostrum 5 supercomputer, jointly funded by EuroHPC JU, Portugal, Turkey and Spain. This work was supported by the HE EuPRAXIA-PP under grant agreement No. 101079773. CB acknowledges the support of the Portuguese Science Foundation (FCT) Grant No. PRT/BD/152270/2021 (DOI: 10.54499/PRT/BD/152270/2021), RA FCT Grant No. UI/BD/154677/2022 (DOI: 10.54499/UI/BD/154677/2022), BM FCT Grant No. PD/BD/150409/2019, and TS FCT IPFN-CEEC-INST-LA3/IST-ID.
\end{acknowledgments}
\appendix*                          
\section{End Matter}
\label{app:PIC}
\emph{Appendix 1: Tailored Density Profile}---\ The ideal plasma-density taper that preserves phase synchronism between the accelerated electrons and the longitudinal wakefield can be obtained by requiring the instantaneous particle velocity $v(z)$ to coincide with the phase velocity $v_w(z)$ of the $n_{w}$-th accelerating bucket throughout the interaction length.

The on-axis accelerating field is proportional to the square root of the plasma density and it can be written as
\setcounter{equation}{0}
\begin{equation}
E_z(z)=\alpha_0\,\frac{m_e\,\omega_{p0} c}{e}\sqrt{\frac{n(z)}{n_0}},
\label{eq:Ez}
\end{equation}
where $\alpha_0$ is the normalized field amplitude. The longitudinal momentum of a test particle of charge $q$ and mass $m_p$ evolves according to
\begin{equation}
\frac{dp_z}{dt}=m_p \frac{{d}[\gamma(z) v(z)]}{{d}t} = q\,E_z(z).
\label{eq:ap2}
\end{equation}
The spatio-temporal driver's peak intensity that is responsible for the wakefield excitation moves at $v_f$. Thus, we replace the time derivative by the corresponding propagation variable of $z=v_f \,t$, leading to
\begin{equation}
\frac{{d}}{{d}z}\bigl[\gamma(z)v(z)\bigr]
      =\alpha_0\,\frac{q}{e}\,\frac{m_e}{m_p}\,\frac{\omega_{p0}c}{v_f}
      \sqrt{\frac{n(z)}{n_0}}.
\label{eq:dgammaU}
\end{equation}
Equation (\eqref{eq:dgammaU}) constitutes the first element of the coupled system that will lead us to our ideal density profile.

The second equation can be derived by noting that each accelerating bucket has an approximate size of $\sim \lambda_p$, the axial position of the center of the $n_{w}$-th accelerating cavity with respect to the driver peak is
\begin{equation}
\xi_w(t)=v_f\, t - n_w\lambda_p\bigl[z(t)\bigr], \qquad
\lambda_p(z)=\frac{2\pi v_f}{\omega_p(z)},
\end{equation}
with $\omega_p(z)=\sqrt{n(z)e^2/(m_e\varepsilon_0)}$.
Differentiation with respect to time gives the phase velocity of the wake
\begin{align}
 v_w & = \frac{{d}\xi_w}{{d}t}=v_f - n_w v_f\frac{{d}\lambda_p}{{d}z} \nonumber \\
        & = v_f-\frac{2\pi n_w v_f^2}{\omega_{p0}}\frac{{d}}{{d}z}
            \sqrt{\frac{n_0}{n(z)}}.
\label{eq:vw}
\end{align}
To obtain the ideal density profile that keeps our particles phase-locked to the ideal acceleration position, we impose that $v_w = v$ when solving Eqs. \eqref{eq:dgammaU} and \eqref{eq:vw}. \\

\emph{Appendix 2: PIC Simulation Details}---All simulations were carried out with the fully relativistic PIC code
\textsc{OSIRIS}~\cite{2002Fonseca} using an azimuthal mode decomposition algorithm with near cylindrically symmetric geometry ($r,z$–$\varphi$), where fields and currents are expanded in azimuthal modes \cite{lifschitz2009particle,davidson2015implementation}. The simulations presented here use the cylindrically symmetric mode ($m=0$) and $m=1$, which is enough to capture the laser and wake structure at reduced computational cost. For a nominal plasma density of
$n_0 = 4\times10^{18}\,\mathrm{cm^{-3}}$ we have
$c/\omega_p = \SI{2.55}{\micro\meter}$.

The simulation was performed in a window co-moving at the speed of light ($\xi = z - ct$) of dimensions  
$\ell_{\xi}=\SI{344}{\micro\meter}$ in the propagation direction and  
$\ell_{r}=\SI{51}{\micro\meter}$ radially, corresponding to  
$L_{\xi}=135\,(c/\omega_p)$ and $L_{r}=20\,(c/\omega_p)$.  
The grid had $27\,000 \times 400$ cells  
($\Delta\xi = 0.005\,k_p^{-1}$, $\Delta r = 0.05\,k_p^{-1}$);  
the time step $\Delta t = 3.5\times10^{-3}\,\omega_p^{-1}$  
($0.030\,\text{fs}$) satisfied the Courant condition.  
The simulations had open boundary conditions for outward fields at large $r$ and absorbing conditions for particles. Dynamic load balancing was enabled every 100 iterations.

The driver was a space–time \textit{flying-focus} wave packet centered at $\lambda_0 = \SI{0.8}{\micro\meter}$ ($\omega_0/\omega_p = 20$), with normalized vector potential $a_0 = 3$ for the case shown in Figure~\ref{fig:1} and $a_0 = 0.8$ in all the other cases, spot size $w_0 = 1.4\,k_p^{-1}$ ($\SI{3.56}{\micro\meter}$), and an envelope length $L_{\mathrm{env}} = 108\,k_p^{-1}$ ($\SI{275}{\micro\meter}$).
Its intensity peak advanced rigidly at the prescribed focus velocity
$v_f$ (Fig.~\ref{fig:1}), while the envelope propagated at $c$.

Witness beams were initialized with flat-top shapes with length and radius $\sigma_z = 75\,k_p^{-1}$ ($\SI{0.2}{\milli\meter}$),  
$\sigma_r = 0.26\,k_p^{-1}$ ($\SI{0.5}{\micro\meter}$), respectively, peak density $n_b = 10^{-3} n_0$ and particles of mass $m_\pi/m_e = 264$, $m_\mu/m_e = 207$, $m_{\mu/2}/m_e = 100$, and $m_{\mu/4}/m_e = 50$. The initial velocity of the particle was initialized for different values depending on the case considered. We performed a comparison of our model [Eq. \eqref{eq:denwake} with simulations up to beam densities $n_b =  n_0$, and found excellent agreement, thus proving that beam-loading effects are not an issue for beam densities on the order of the plasma density.

In simulations, the tailored plasma pro\-fi\-les $n(z)$ fol\-lo\-wed the analytical prescription of Eqs.\, (1) that maintains phase-locking while the beam accelerates.
\\

%
\bibliography{bibliography}
\end{document}